\documentclass[aps,prd,twocolumn,amsmath,amssymb,showpacs]{revtex4-1}
\usepackage{graphicx}
\usepackage{dcolumn}
\usepackage{bm}

\usepackage{latexsym}
\usepackage{amsthm}
\usepackage{braket}

\usepackage[top = 1. in,bottom = 1. in, left = 1 in, right=1 in]{geometry}

 \newcommand{\arXiv}[1]{\href{http://www.arXiv.org/abs/#1}{#1}}
\usepackage[colorlinks=true, linkcolor=blue, citecolor=blue, bookmarks=true]{hyperref}

\makeatletter
\renewcommand\section{\@startsection {section}{1}{\z@}%
                  {-3.5ex \@plus -1ex \@minus -.2ex}
                  {2.3ex \@plus.2ex}%
                  {\normalfont\large\bfseries}}
\renewcommand\subsection{\@startsection{subsection}{2}{\z@}%
                   {-3.25ex\@plus -1ex \@minus -.2ex}%
                   {1.5ex \@plus .2ex}%
                   {\normalfont\bfseries}}
\makeatother

\newcommand{\be}{\begin{equation}}
\newcommand{\ee}{\end{equation}}
\newcommand{\beq}{\begin{equation}}
\newcommand{\eeq}{\end{equation}}
\newcommand{\bea}{\begin{eqnarray}}
\newcommand{\eea}{\end{eqnarray}}

\newcommand{\eq}[1]{(\ref{#1})}
\newcommand{\fig}[1]{Fig.~\ref{#1}}

\begin{document}

\title{Holography and  thermalization in optical pump-probe spectroscopy}
\author {A.~Bagrov,$^a$ B.~Craps,$^{b}$ F.~Galli,$^{c}$ V.~Ker\"anen,$^d$ E.~Keski-Vakkuri,$^d$ J.~Zaanen$^e$\vspace{2mm}}

\affiliation{ $^a$Institute for Molecules and Materials, Radboud University, Nijmegen, The Netherlands \\
$^b$Theoretische Natuurkunde, Vrije Universiteit Brussel and The International Solvay Institutes, Brussels, Belgium \\
$^c$Perimeter Institute for Theoretical Physics, Waterloo, Ontario, Canada\\
$^d$Department of Physics, University of Helsinki, Helsinki, Finland\\
$^e$Instituut-Lorentz for Theoretical Physics, Universiteit Leiden, Leiden, The Netherlands}

\begin{abstract}

Using holography, we model experiments in which a 2+1D strange metal is pumped by a laser pulse into a highly excited state, after which the time evolution of the optical conductivity is probed. We consider a finite-density state with mildly broken translation invariance and excite it by oscillating electric field pulses. At zero density, the optical conductivity would assume its thermalized value immediately after the pumping has ended. At finite density, pulses with significant DC components give rise to slow exponential relaxation, governed by a vector quasinormal mode. In contrast, for high-frequency pulses the amplitude of the quasinormal mode is strongly suppressed, so that the optical conductivity assumes its thermalized value effectively instantaneously. This surprising prediction may provide a stimulus for taking up the challenge to realize these experiments in the laboratory. Such experiments would test a crucial open question faced by applied holography: Are its predictions artefacts of the large $N$ limit or do they enjoy sufficient UV independence to hold at least qualitatively in real-world systems?

\end{abstract}

\maketitle

\section{Introduction}   
 The quantum non-equilibrium physics of strongly interacting many body systems \cite{polkovnikovetal}
  is a largely unexplored frontier. It is driven in part by experimental progress in ultrarelativistic heavy-ion collisions \cite{Adams}, cold atom systems \cite{Bloch} and quantum simulators \cite{Lanyon}, as well as by emerging ``ultrafast" techniques in condensed matter physics \cite{Orenstein}.  A classic example of quantum non-equilibrium experiments are optical pump-probe experiments: a system is excited by a very short and intense coherent electromagnetic pulse, and the time evolution of the system after the pulse is monitored by a linear response probe \cite{DalConte,Giannettietal,Freericks}.  
 
 On the theoretical side, one is dealing in this non-equilibrium setting with highly excited states of  strongly interacting quantum many body systems  characterized by dense many body
 entanglement. Given its quantum complexity a quantum computer is required to compute with confidence. However, evidence has been accumulating that 
 holographic (or gauge/gravity) duality
 may catch generic features of such systems. Holography translates the physics of quantum many body systems into a dual gravitational problem in a space-time with an extra 
dimension (e.g., see the textbooks \cite{Ammon:2015wua,Zaanenetalbook}). It is actually ideally suited to study non-equilibrium physics, which is often mapped into tractable 
non-stationary general relativity (GR) problems, demonstrating considerable success in near equilibrium circumstances \cite{Zaanenetalbook},  e.g.\ in the form of  the ``minimal viscosity" \cite{Kovtun:2004de} and ``Planckian  dissipation" \cite{Hartnoll14}. 

Holography rests on the matrix large $N$ limit governing the ultraviolet (UV) of the boundary field theory and it is a priori unclear when one is dealing with pathologies associated
with this limit. The pump-probe set-up is a natural theatre to study these questions in far out-of-equilbrium circumstances. 
One type of pump-probe experiments consists in pumping the (thermal) vacuum of a zero-density quantum critical system described by a conformal field theory (CFT) out of equilibrium by a  homogeneous  and unitary pulse of transversal electrical field. This is dual to the Vaidya metric \cite{Horowitz:2013mia}, a  non-stationary analytical GR solution describing an infalling shell of null dust collapsing into a black hole. In this case, one-point functions as well as retarded two-point functions with both operators inserted after the time of energy injection coincide  with their thermal equilibrium values. This ``instantaneous thermalization" \cite{Bhattacharyya:2009uu} of one-point functions and retarded two-point functions is obvious from the bulk  perspective: by causality, if a perturbation is applied after the quench, it is only sensitive to the region of spacetime outside the shell, where the bulk geometry coincides with the black hole geometry that describes thermal equilibrium.

Such instantaneous thermalization is incomprehensible viewed from a (semi) classical perspective. However, we are dealing with quantum thermalization:  given the ``eigenstate thermalization hypothesis" insight \cite{Deutsch, Srednicki},  there is a priori no speed limit associated with the rate that a local observer loses track of the flow of information in the exponentially large many-body Hilbert space. In contrast, nonlocal observables including the bipartite entanglement entropy do reveal a nonzero thermalization time in the Vaidya set-up \cite{Hubeny:2007xt, AbajoArrastia:2010yt, Albash:2010mv, Balasubramanian:2010ce,Balasubramanian:2011ur}.  A recent study addressed these matters directly in a  zero-density 1+1D CFT relating the Vaidya shell to structure realized in the large central charge limit \cite{Anous:2016kss, Anous:2017tza}, suggesting that the instantaneous thermalization might be a large $N$ artefact. Another recent study addressed thermalization of fermionic Green's functions after a quench in the Sachdev-Ye-Kitaev model \cite{Eberlein:2017wah}, which is believed to be a quantum system with a holographic dual. For $q$-fermion interactions with $q\rightarrow\infty$, instantaneous thermalization was also found, consistent with an AdS$_2$-Vaidya dual.

Holographic strange metals behave very differently from zero-density CFTs, revealing suggestive resemblances with the  
strange metals found in high $T_c$ superconductors and related systems \cite{Keimer}. There are reasons to believe that this is rooted in dense vacuum entanglement due to fermion signs \cite{Zaanenetalbook, Grover}. 
Here we will present a holographic construction that describes a 2+1D strange metal that is pumped by a short transversal electrical field pulse in a spatially homogeneous way into a highly excited state, after which the time evolution of the optical conductivity is probed.  We mimic quite closely state-of-the-art  pump-probe experiments \cite{DalConte,Giannettietal}. We specifically consider the simple ``RN strange metal'', dual to a charged Reissner-Nordstr\"om (RN) black brane, pumped by a  homogeneous transversal electrical field pulse that is oscillating in time (like a laser pulse) and  
characterized by a mean frequency $\omega_P$ (\fig{figure_1}).   We subsequently compute the probe optical conductivity at times after the pulse, as is the case in   pump-probe experiments.   To bring the optical conductivity alive we mimic the breaking of translational invariance  by incorporating momentum relaxation through linear axions \cite{Andrade:2013gsa} (see \cite{Vegh:2013sk,Blake:2013owa,Donos:2013eha} for related models).  

The excited states of the strange metal are very different from those of the zero density CFT and by varying the pulse frequency $\omega_P$ relative to  the chemical potential $\mu$ we can study numerically
how this affects the thermalization process. {\em The outcome is that in all circumstances the thermalization continues to be 
instantaneous} (\fig{sigmas}a, \ref{sigma_vs_t}), except when the pulse contains a static electrical field component. The latter would set the charges in uniform motion, delaying thermalization by (half of) the final state equilibrium momentum relaxation time (\fig{sigmas}b, \ref{sigma_vs_t}). In analogy with the minimal viscosity \cite{Kovtun:2004de}, we conjecture that 
 this ultrafast quantum thermalization  may be general for large classes of theories with gravitational duals. We also address the question of how to test these predictions in the laboratory. 
 
Our work proposes a novel way to test the crucial question whether holography describes real-world physical systems. Gravitational computations are valid in a matrix large $N$ limit, of which no realization is known in nature. An important reason why holography has nevertheless been applied to condensed matter systems is the expectation that the strange metal regime of finite density CFTs enjoys strong emergence, i.e.\ that the relevant infrared physics is largely independent of details of the UV theory (such as large $N$). Because of the sign problem, it is impossible today to decide on theoretical grounds whether this expectation is justified, i.e.\ whether the results extend at least qualitatively to finite $N$. We propose to address this question experimentally using pump-probe spectroscopy of the strange metals in e.g.\ the cuprates. In fact, there is already evidence for such UV independence, in that the minimal viscosity arising in large $N$ holographic computations appears to be observed in the quark gluon plasma (which has $N=3$) and even in the unitary Fermi gas of cold atom physics (whose UV is completely disconnected from matrix field theory); see e.g.\ \cite{Schafer:2009dj, Wlazlowski:2012jb}. Similarly, the extremely short time for hydrodynamic behavior to set in after a collison of heavy ions appears to be reproduced by holographic computations \cite{Chesler:2008hg} related to our instantaneous thermalization, providing additional motivation to study this phenomenon in the condensed matter laboratory using the simplest circumstances available, namely optical pump-probe spectroscopy. 
 
 
\section{Setup} 
Our holographic setup is the standard minimal Einstein-Maxwell action \cite{Zaanenetalbook} extended by linear axion scalar fields  $\phi_I$ \cite{Andrade:2013gsa}, 
\beq
S = \frac{1}{2\kappa_4^2}\int d^4x\,\sqrt{-g}\Big[R - 2\Lambda - \frac{1}{2}\sum_{I = 1}^2(\partial\phi_I)^2
- \frac{1}{4} F^2\Big]\ .
\eeq
The RN strange metal is dual to a charged black brane in the bulk with non-vanishing gauge and scalar field configurations,
\begin{figure}[h!]
\begin{center}
 		\includegraphics[width=\columnwidth]{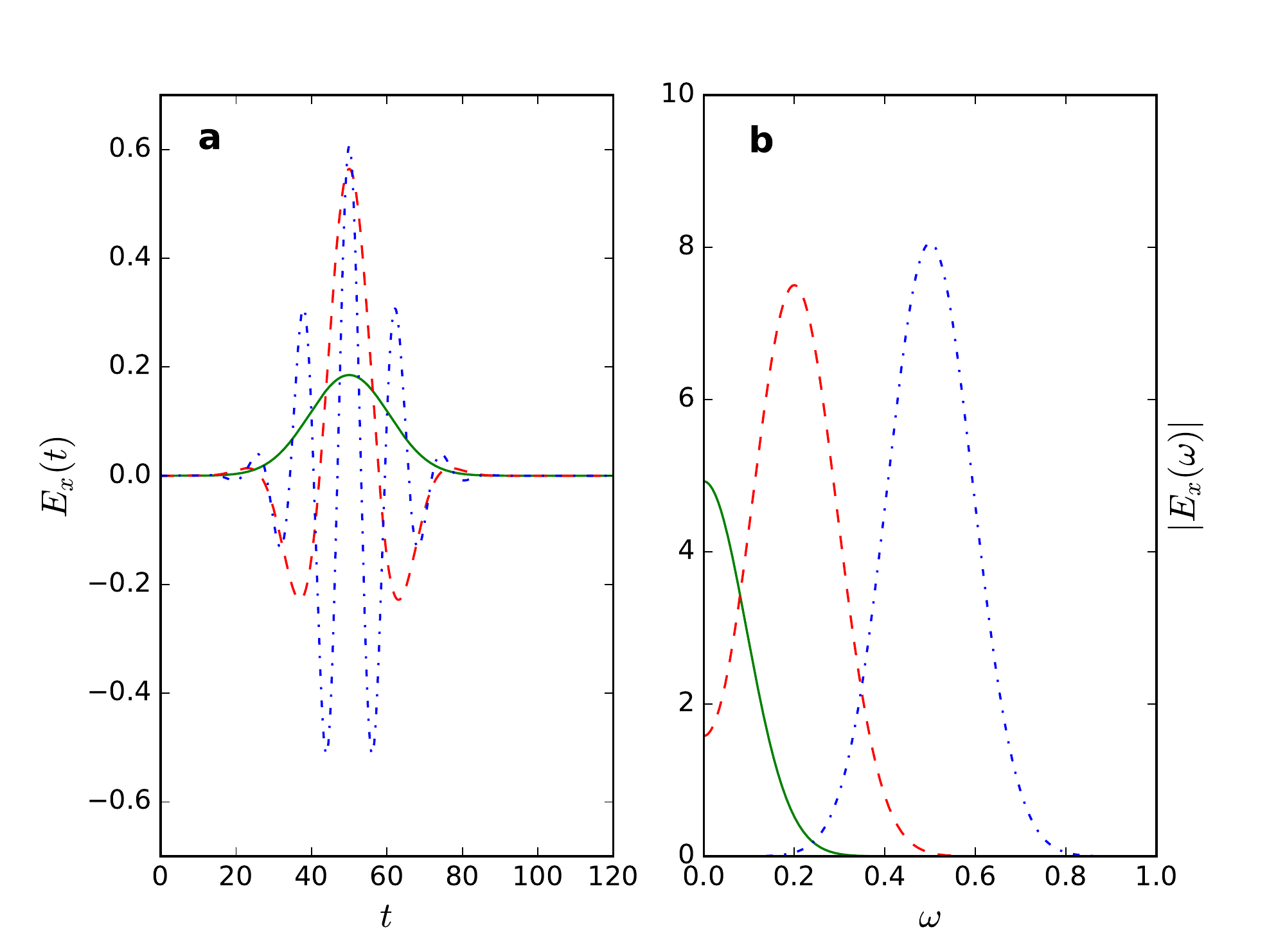}
  	\end{center}
  	\caption{\small{Pump electric field profiles in the boundary both in the time- and frequency domain (left- and right panels) measured in units of the initial chemical potential $\mu=1$. These are characterized by mean frequencies $\omega_{P} = (0, 0.2, 0.5)$ corresponding with the  solid, dashed and dot-dashed lines, respectively.}}
 	\label{figure_1}
 \end{figure}
\begin{align}\label{eq:equilibrium_solution}
&ds^2 = \frac{1}{z^2}\Big( - f dt^2 + \frac{dz^2}{f} + dx^2 + dy^2\Big)\ ,
\\
&f  = 1 - \frac{1}{2}k^2 z^2 - m z^3 +\frac{1}{4}\rho^2 z^4\ ,
\\
&A = (\rho z-\mu )dt \ , \\
&\phi_1 = k x \, ,\qquad \phi_2 = k y \ , \label{eq:linearscalar}
\end{align}
where $x$ and $y$ are coordinates in the field theory directions.  The axions cause momentum dissipation in the boundary, parametrised by  $k$. This mimics the effects of  explicit translational symmetry breaking, which is much harder to implement \cite{Horowitz:2012ky}. It is however well understood \cite{Zaanenetalbook} that in the regime of interest  (zero momentum transport, weak potentials) the axions are representative and $\ell_{\rm mfp}\equiv1/k$ can be interpreted as an elastic mean free path. According to standard holography, the entropy density is $s=4\pi/z_0^2$, with $z_0$ the horizon location, $f(z_0)=0$. The charge density $\rho$ is related to the chemical potential by $\mu=\rho z_0$. Formulae for the temperature $T$ and (relativistic) energy density $\epsilon$ are provided in Appendix~\ref{app:equilibrium}.

To study the effect of a time-dependent external electric field $E_x(t)$ in the $x$ direction, we consider the ansatz \cite{Withers:2016lft}
\begin{align}
&ds^2 =- F_z( z,v)dv^2 -  \frac{2}{z^2} dv dz + 2F_x(z, v) dx dv \nonumber
\\
&\quad\ \  + \Sigma(z, v)^2(e^{-B(z, v)}dx^2 + e^{B(z, v)}dy^2) \ .
\\
&A= (E_x(v) x + a_v(z, v))dv + a_x(z, v)dx \, ,
\\
&\phi_1 = k x + \Phi(z, v)\, , \quad \phi_2 = k y\, 
\end{align}
and solve the bulk equations of motion numerically, subject to appropriate boundary conditions.  The coordinate $v$ coincides with the field theory time $t$ at the boundary $z=0$. 
We use electric field profiles 
\begin{equation}
E_x(t) = A \cos(\omega_{P} t) e^{-\frac{(t - t_0)^2}{(\Delta t)^2}} \frac{1 - \tanh \frac{t - t_0 - 3\Delta t}{\delta}}{2},
\end{equation}
displayed in \fig{figure_1}. 
We consider central frequencies $\omega_{P} = (0, 0.2, 0.5)$, modulated by a Gaussian envelope of width $\Delta t = 15$ centered at $t_0=50$ and cut off by a smoothed step function ($\delta = 0.01$) at $t_{\rm end}\equiv t_0+3\Delta t =95$ (from which time onwards we consider the pumping to have finished).
The amplitude $A \approx (0.19, 0.56, 0.61)$ is tuned for each $\omega_P$ such that the temperature increases from $T_I=0.2$ to $T_F=0.3$. We set the initial chemical potential $\mu=1$, and normalize all other quantities accordingly.

At zero charge density, the above system is immediately solved by the Vaidya metric \cite{Horowitz:2013mia}.
Momentum dissipation can be easily incorporated, continuing to give rise to a Vaidya-like bulk geometry \cite{Bardoux:2012aw} (see also \cite{Withers:2016lft}).  As soon as a nonzero charge density is introduced, 
the gravity solution becomes more complicated, and one has to resort to numerical general relativity \cite{Withers:2016lft}. The physical mechanism behind this in the boundary theory is that the electric field pulse will set the charges in motion, which induces corresponding contributions to the energy-momentum tensor; in the bulk theory, this corresponds to exciting additional metric components. As described in Appendix~\ref{app:numerical} (and in more detail in \cite{preparation}), the resulting system can be solved using established numerical methods described in e.g. \cite{Chesler:2008hg, Heller:2013oxa, Chesler:2013lia, Ecker:2015kna}.


\section{Probe conductivity} We are interested in the time-dependent optical conductivity. In terms of a spatially homogeneous probe electric field $\delta E_i(t)$ ($i$ is the spatial coordinate, we later set $i = x$) and the resulting current perturbation $\delta \langle J_i(t)\rangle$, the differential conductivity $\sigma_{ij}(t, t')$ is defined through
\beq
\delta \langle J_i(t)\rangle = \int_{t_i}^{t} dt' \sigma_{ij}(t, t') \delta E_j(t')  \ ,\label{eq:difcond}
\eeq
where $t_i$ is the time when the probe experiment starts, which we take to be the time $t_{\rm end}=95$ at which the pumping ends. 
The differential conductivity can be computed by choosing the probing electric field to be  
$\delta E_j(t) = \varepsilon \delta(t - \bar t)\delta_{jx}$,
with $\varepsilon$  being small. (In our numerical computations, we replace the delta function by a narrow Gaussian.) Substituting this into (\ref{eq:difcond}), we obtain
$\sigma_{xx}(t, \bar t) = \delta\langle J_x(t)\rangle/\varepsilon$. For a time-dependent state, there is no unique definition of a frequency space conductivity. 
Following \cite{Lenarcic2014}, we use the definition
\beq
\sigma(\omega, t) = \int_t^{t_m}dt' e^{i\omega(t' - t)}\sigma_{xx}(t', t)\ ,\label{eq:cond}
\eeq
where $t_m$ is  determined by the duration of the probe.   
(We will use $t_m =3446.18$ for the numerical evaluation of  $\sigma(\omega, t)$.)

\section{Results}  
\begin{figure}[h!]
\begin{center}
 		\includegraphics[width=\columnwidth]{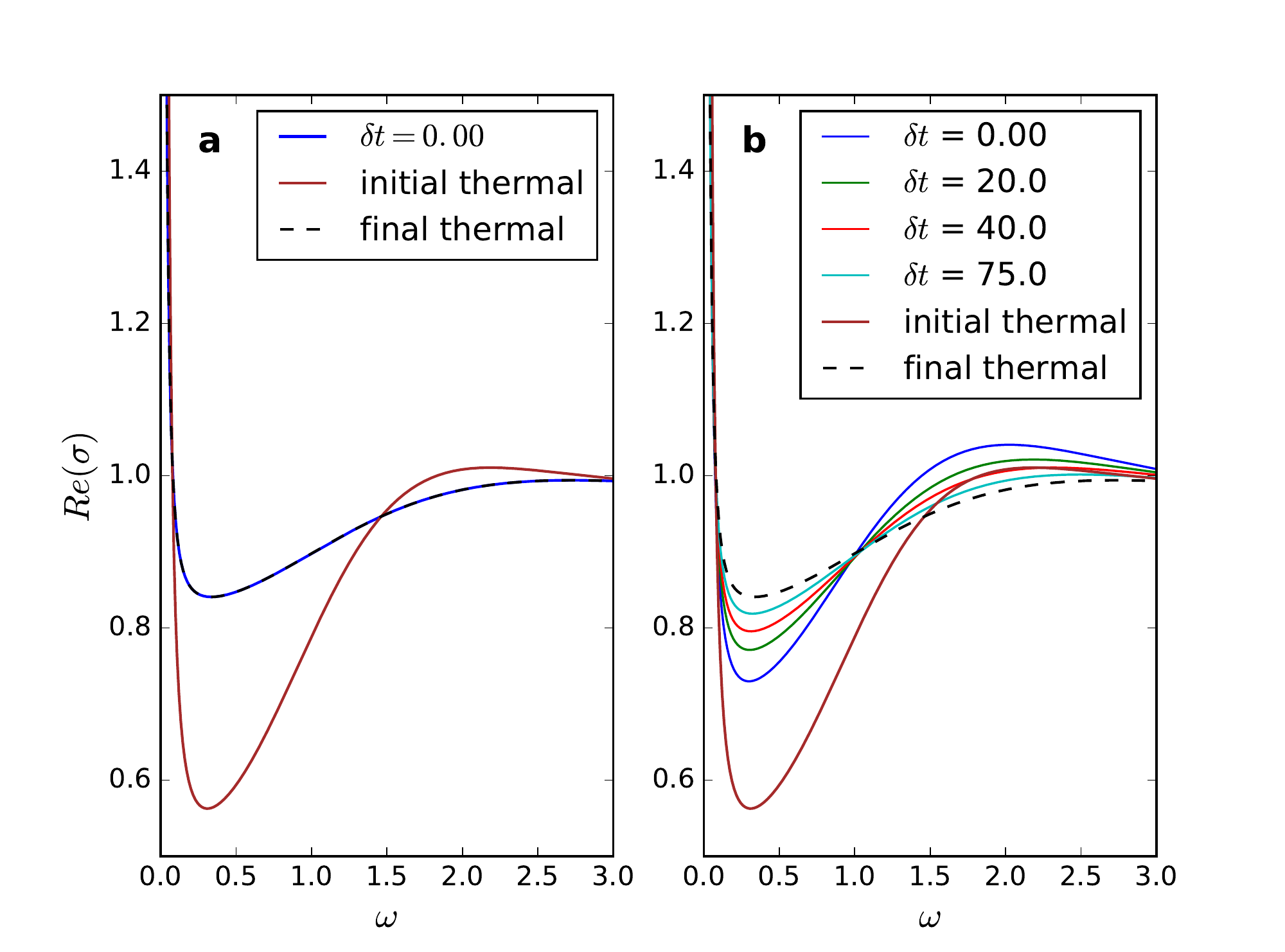}
  	\end{center}
  	\caption{\small{Representative examples of the real part of (probe) conductivities as functions of $\omega$. (a)  The real part of the optical conductivity for the initial equilibrium state (solid brown) and for the excited state right after the pump pulse with $\omega_P = 0.5$ has ended (solid blue). For this pulse, the conductivity at all times after the system has been excited essentially coincides with the equilibrium result at the final temperature $T_F$ (dashed black). (b) The time-dependence of the same system with the only difference that it is now excited by a pump electrical field with vanishing $\omega_P$. The time $\delta t = t - t_{\rm end}$, measured after the termination of the pump pulse, and the parameters $T_{I} = 0.2$, $k = 0.2$ and $T_{F} = 0.3$ are all expressed in units where the initial chemical potential $\mu=1$.}}
 	\label{sigmas}
 \end{figure}

The optical conductivity is a convenient probe to keep track of the time-dependence of the equilibration process (see \fig{sigmas}a).  For weak momentum relaxation, 
it is characterized in equilibrium by a Drude peak at low energy  with weight  set by the charge density and width by the momentum relaxation rate $1/ \tau_{Q}$.
At energy $\simeq \mu$ it turns into the energy- and temperature independent  response associated with the 2+1D CFT
(see e.g. \cite{Zaanenetalbook,Horowitz:2012ky}). 
The momentum relaxation time for the RN strange metal is well known \cite{Davison:2013jba}  in terms of the (relativistic) energy density, entropy density and momentum dissipation parameter of the boundary theory:
$\tau_{Q} =  6\pi \epsilon / ( sk^2)$. 

Our main result is that under the sole condition that the pulse does not contain a zero-frequency component (\fig{figure_1})
 the optical conductivity measured at {\em all} times after the pulse is {\em 
identical} to the one characterizing the equilibrium system at the final temperature $T_F$. (More precisely, the condition is that the Fourier transform of the pulse should not have support at the lowest quasinormal mode (QNM) frequency $\omega_*$, which is purely imaginary \cite{preparation}.) 
In \fig{sigmas}a we show the outcome for $\omega_P = \mu/2$; notice that 
we can easily accomplish a final temperature $T_F$ which is substantially higher than the initial temperature $T_{I}$ by tuning the ``fluency" $A$ of the pulse (see \cite{preparation} for more details).   Regardless whether we excite the strange metal states ($\omega_P < \mu$) or those of the zero density CFT ($\omega_P > \mu$),  we find that the instantaneous thermalization is ubiquitous. 

\begin{figure}[h!]
\begin{center}
 		\includegraphics[width=\columnwidth]{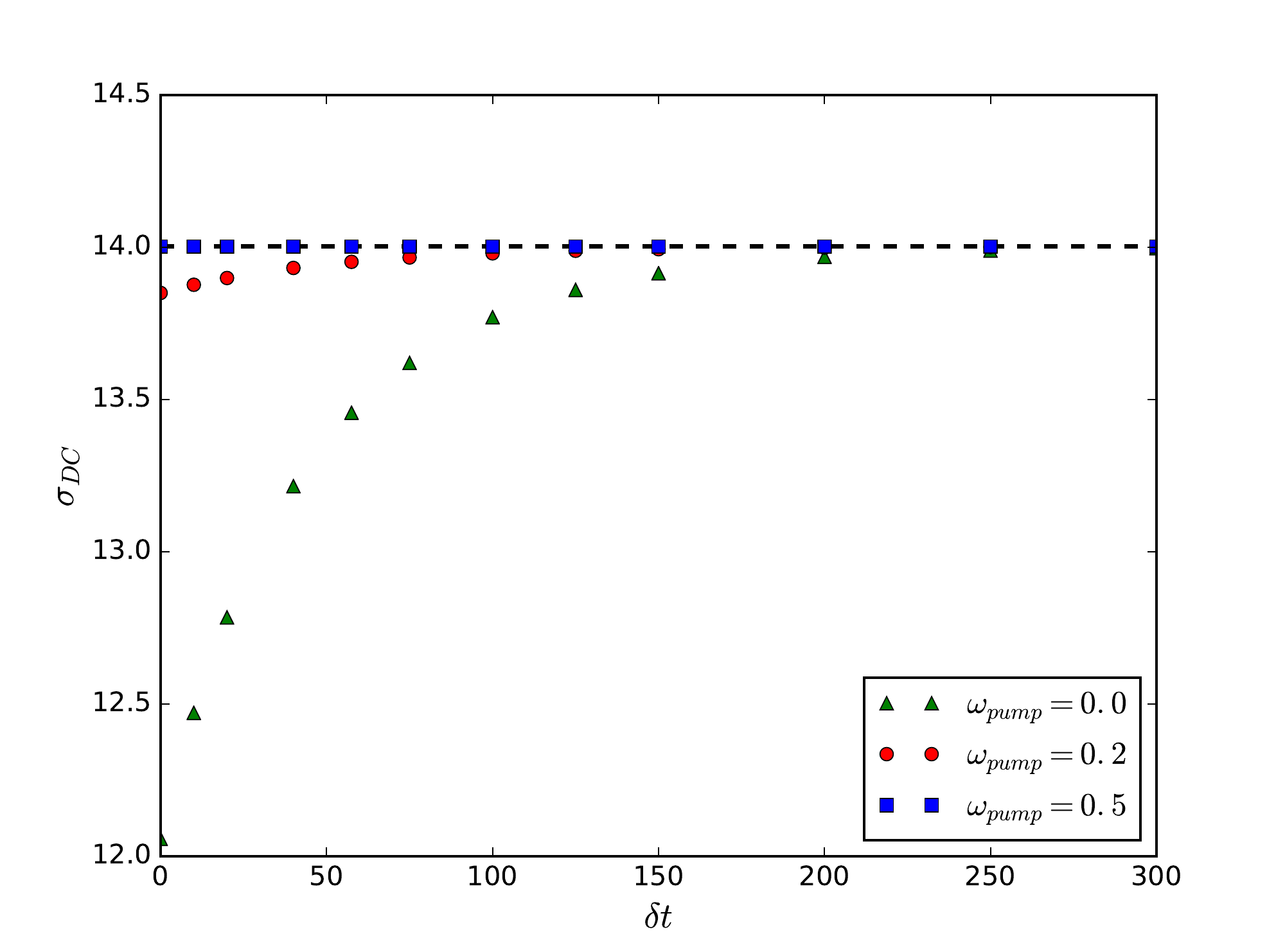}
  	\end{center}
  	\caption{\small{The time-dependence of the DC conductivities for the same parameters as in \fig{sigmas}: The green triangles, the red circles and blue boxes correspond to 
	the pulses of \fig{figure_1} with $\omega_p = 0, 0.2$ and $0.5$, respectively. The dashed line corresponds to the equilibrium DC conductivity at temperature $T_F$.   The time scale is half the momentum relaxation time, while the magnitude is associated with the zero frequency components of the pulse. }}
	\label{sigma_vs_t}
 \end{figure}

The situation changes drastically when the pulse probe contains a zero frequency component. A typical case is shown in \fig{sigmas}b:
we used here the same setup as for \fig{sigmas}a with the
only difference that we now excite the system with a $\omega_P = 0$ pulse. The nature of the time-dependence becomes transparent focussing in on 
the DC conductivity as a function of the time since the pumping ended, $\delta t \equiv t - t_{\rm end}$  (\fig{sigma_vs_t}).  This is fitted precisely with the simple form  
$\sigma_{DC} (\delta t)  = \sigma_{DC}^{th} - C  \exp{(- \delta t/\tau_{\rm therm} )}$, where $\sigma_{DC}^{th}$
is the DC conductivity in the final equilibrium state with temperature $T=T_F$. This form 
indicates that the full time evolution is governed by just a single thermalization  time $\tau_{\rm therm}$. 
We find that {\em the thermalization time is precisely half the equilibrium momentum relaxation time at $T_F$}: $\tau_{\rm therm} = \tau_{Q}/2$. This is due to the fact that the metric components induced by the momentum of the charges appear squared in computation of the optical conductivity \cite{preparation}. 
The amplitude of this decaying exponential is governed by the shape of the pump pulse: for a pulse with central frequency $\omega_P$ and Gaussian envelope with width $\Delta t$, it is
$C \propto  \exp{ (- (\Delta t)^2 \omega^2_P/2)}$, corresponding to the magnitude squared of the zero frequency pulse component (\fig{sigma_vs_t}). (More generally, one should evaluate the Fourier transform of the pump pulse at the lowest QNM frequency, which for Gaussian envelopes gives the same suppression factor \cite{preparation}.)
 
At first sight this looks confusing: how can energy relaxation  be governed by  a momentum relaxation time? 
There is however a simple physical explanation. The zero frequency component in the  pulse corresponds with just a static electrical field that is switched on during the  pulse.  
This will accelerate the finite density system, which will carry a finite total momentum when the pulse is over. 
Momentum relaxation is required for the system to come to a standstill, taking the time $\tau_{Q}$  to turn its kinetic energy into a full heating of the system. Notice that at zero density this delay is  absent since the electrical current decouples from momentum.  

As we substantiate in Appendix~\ref{app:relating}, this has an elegant bulk interpretation. In the linearized approximation, the decay towards the late-time static black hole is typically governed by the imaginary part of the lowest QNM frequency $\omega_*$, namely $\tau_{Q} =  -1/{\rm Im}(\omega_*)$. For pump pulses with a static component, we find that the corresponding exponential decay holds remarkably well even in the fully nonlinear regime. (Surprisingly good agreement with a linearized approximation was emphasized before in the context of holographic models for heavy ion collisions, see e.g.\ \cite{Heller:2013oxa}.)   For pump pulses devoid of a zero frequency component, lack of resonance suppresses the QNM excitation, leading to effectively instantaneous thermalization. This mechanism depends on the fact that energy is injected via a spatially homogeneous electric field and on the fact that a black hole is formed, but not on finer details of the geometry. On this basis {\em we conjecture that large classes of field theories with a gravitational dual will exhibit the phenomenon of instantaneous thermalization} when driven out of equilibrium by a pulse of electromagnetic energy. 

Pump electric fields with a significant DC component have been previously investigated in \cite{Withers:2016lft}, where one-point functions of electric and heat currents and the QNMs governing their decay were studied in detail, also for faster momentum relaxation (for which qualitative differences were found). In the present work, we focus on the regime of slow momentum relaxation, which we consider the most relevant regime for describing strange metals, and consider also oscillating pulses to mimick laser light. Moreover, we compute the retarded two-point functions corresponding to conductivities, which do not follow straightforwardly from one-point functions via large $N$ factorization because factorized contributions cancel in commutators. The main novelty, however, is the translation of these holographic findings to experiments that are in principle feasible in condensed matter laboratories.


\section{Discussion}  The pump-probe experiments appear to be an ideal ground to  test the large $N$ ``UV (in)dependence" of holography. Can such instantaneous (or at least, very fast)
quantum thermalization be determined independently? In the field theory one faces the quantum complexity brick wall -- notice that a similar ``instantaneous" response 
was found in a  quantum thermalization problem with polynomial complexity occurring in a charge density wave quantum critical state, giving rise however to an 
instantaneous {\em reactive} (instead of dissipative) response \cite{Freericks}. We present it as a challenge to find out whether the holographic  predictions
 can be tested in the laboratory.
This is not straightforward. The cuprate strange metals are a natural theatre as candidate ``maximally" entangled systems. 
One practical difficulty is that high frequency components in the pulse should be avoided since these will probe the microscopic 
``shake the spins and holes" processes \cite{Giannettietal}.  The pulse energy should be low enough ($<$ 0.5 eV) to make sure that one excites only 
the strongly entangled excitations of the strange metal.  Given the
small mass of the electron, the characteristic time
scales of electronic relaxation processes are by default
in the femtosecond regime, so one should be able to turn off the pump pulse very fast and the probe pulse
needs very good time resolution in order to be sensitive
to anomalously short thermalization times.


\begin{acknowledgements}
We thank T.~Andrade, C.~Ecker, A.~Ficnar, B.~Gouteraux, M.P.~Heller, D.H.~Lee, L.~Rademaker, S.A.~Stricker and D.~Thompson for helpful discussions. 
This research has been supported in part by BELSPO (IAP P7/37), FWO-Vlaanderen (projects G020714N, G044016N and G006918N),
Academy of Finland (grant no 1297472), and the National Science Foundation (grant 
no NSF PHY-1125915). Research at Perimeter Institute is supported by the Government of Canada through Industry Canada and by the Province of Ontario through the Ministry of Research \& Innovation.
\end{acknowledgements}




\appendix

\section{Equilibrium solution}\label{app:equilibrium}
For completeness, we provide expressions for the temperature and energy density of the equilibrium solutions \eq{eq:equilibrium_solution}
\bea
&&  T = \frac{1}{4\pi z_0} \left( 3-\frac{k^2z^2_0}{2}-\frac{\mu^2z^2_0}{4}\right) ,\\
&& \epsilon = 2m = \frac{2}{z^3_0}\left(1-\frac{k^2z^2_0}{2}+\frac{\mu^2z^2_0}{4}\right) \ .
\eea


\section{Numerical method}\label{app:numerical}
Here we provide a brief summary of the numerical method we have used; more details will be provided in \cite{preparation}.
Introducing the dot derivative $\dot{X} \equiv \partial_v X - \frac{z^2}{2} F_z \partial_z X$,
 the equations of motion involving time derivatives turn into  ordinary differential equations for the original fields and their dot derivative fields on a constant $v$ slice.  These are integrated using the 
 Chebyshev pseudospectral approach. The differential equation for $\Sigma$ is non-linear and it is solved using the Newton-Raphson method -- the main technical difference with earlier work.
 The remainder involves then only linear algebra, and the time evolution is obtained using the Runge-Kutta method using the data for the fields an dot derivative fields on a constant $v$ slice. 
This procedure reduces for the  probe conductivity to  linearized equations in the numerically determined  background, such that  the equations of motion turn into simply linear algebraic equations. The numerical calculations are performed using the Python programming language and the SciPy package.


\section{Analytic bulk solution for large-frequency pump pulse}
For large-frequency pump pulses, the bulk solution can be studied analytically. Assume the electric field is of the oscillating form
\beq\label{elec}
E_x(t) = \cos(\omega t)\Omega(t),
\eeq
where $\Omega(t)$ is a smooth function of compact support that varies more slowly than the cosine. We look for solutions in a $1/\omega$ expansion. Inspired by the $\omega$ dependence observed in numerical simulations, we make an ansatz in which the leading contribution to $F_x$ scales like $1/\omega$, while $B$, $\Phi$ and $a_x$ start at order $1/\omega^2$. 
To leading order in $1/\omega$,  the bulk equations of motion are then solved by the Vaidya spacetime
\begin{align}
ds^2 = \frac{1}{z^2}& \Big[  -  \left(1 - \frac{1}{2}k^2 z^2 - M(v) z^3 + \frac{1}{4}\rho^2 z^4 \right) dv^2   \nonumber  \\
& - 2 dv dz + dx^2 + dy^2 \Big] \, ,
 \end{align}
where the mass function is given by
\beq
M(v) = m+\frac{1}{2}\int^v_{-\infty} dv' E_x(v')^2 \label{eq:m(v)}\, ,  
\eeq
 the scalars by \eq{eq:linearscalar} and the leading order gauge field by $A = (\rho z - \mu + x E_x(v)) dv$.
The first subleading correction to the metric, which potentially causes deviations from instantaneous thermalization, is
\beq\label{Fxint}
F_x=\frac13\rho z\int_{-\infty}^v dv'\,E_x(v').
\eeq
For times $v$ during the pump pulse, it follows from \eq{elec} that $F_x$ is indeed of order $1/\omega$. What matters for the probe experiment, however, is the behavior of the metric after the pump pulse has ended. Since $\Omega$ is smooth, $F_x$ at those times is suppressed more strongly than any inverse power of $\omega$, as follows from basic Fourier analysis; for instance, if $\Omega(t)$ were Gaussian with width $\Delta t$, $F_x$ would be suppressed by $\exp(-(\Delta t)^2\omega^2)$. Even though our $1/\omega$ expansion assumed that $\omega$ was the largest scale in the problem, this strong suppression may cause $F_x$ to be small even for reasonably large $\rho$, as suggested by \fig{sigma_vs_t}. 


\section{Relating momentum relaxation and thermalization}\label{app:relating}
In the main text, we claimed that the timescale $\tau_{\rm therm}$ governing the approach to equilibrium of the DC conductivity  is half the timescale $-1/{\rm Im}(\omega_*)$ govering momentum relaxation. \fig{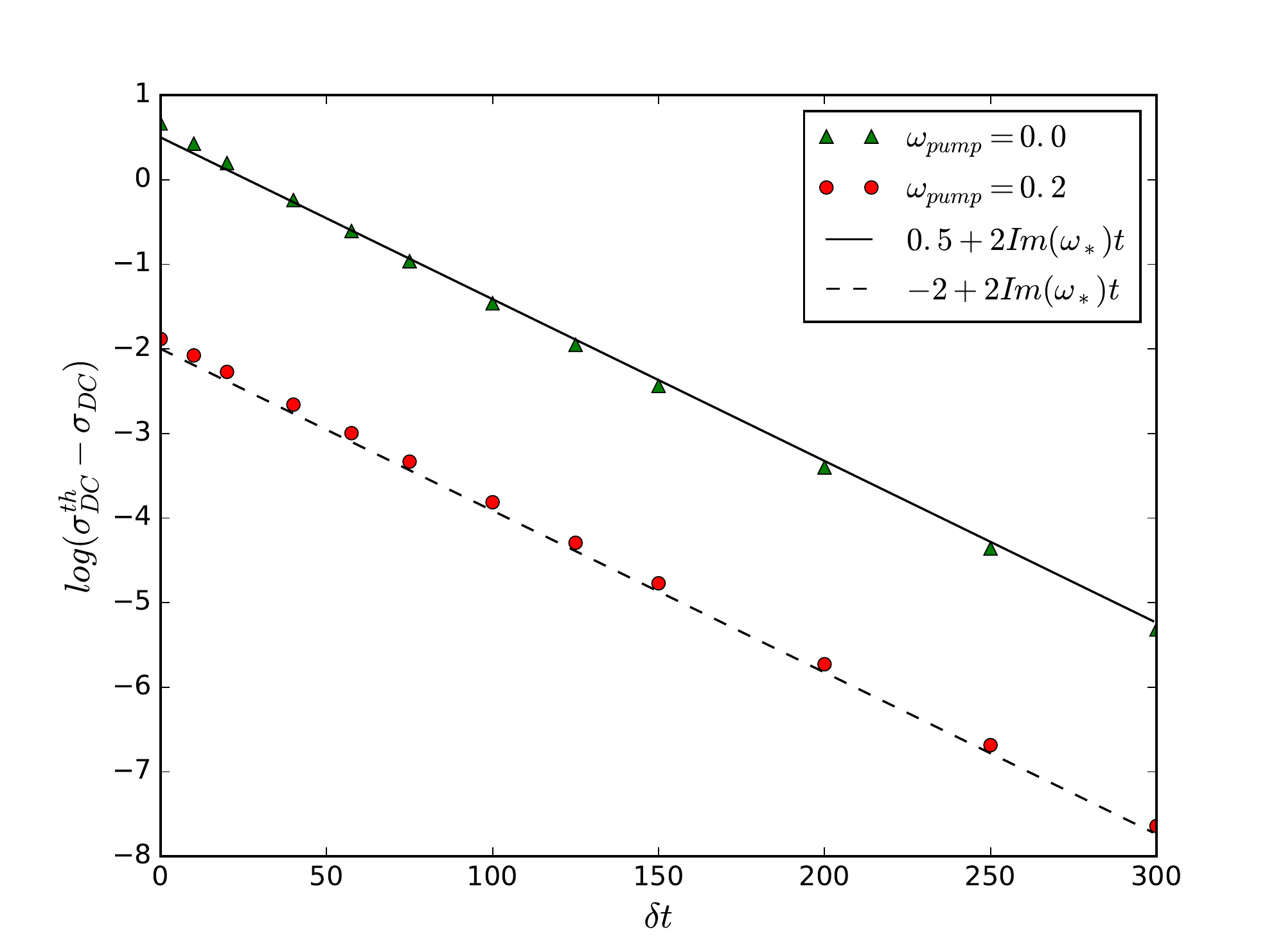} provides numerical evidence in support of this.
 \begin{figure}[h!]
\begin{center}
 		\includegraphics[width=\columnwidth]{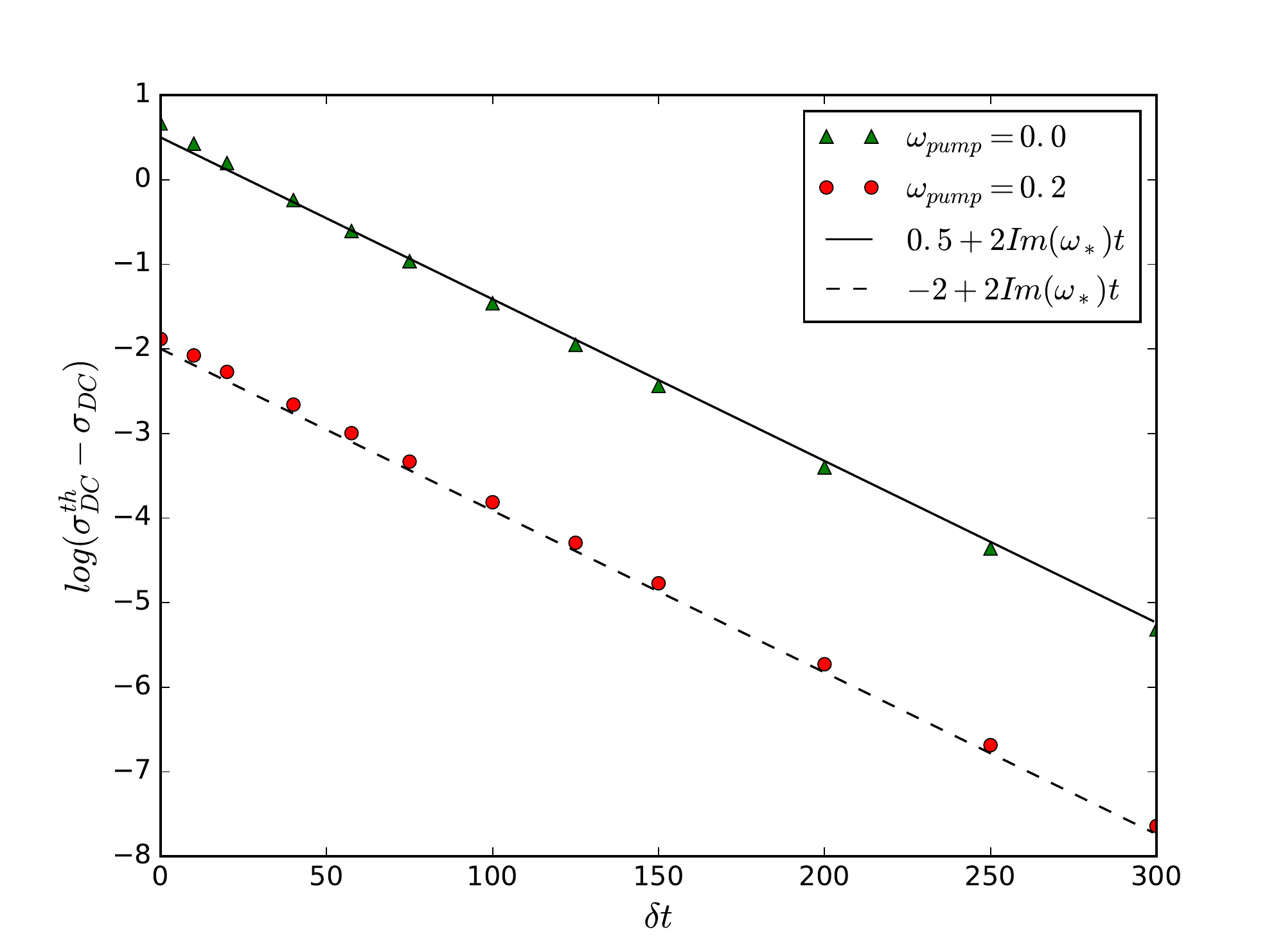}
  	\end{center}
  	\caption{\small{The DC conductivity decays towards equilibrium with a rate consistent with $-2{\rm Im}(\omega_*)$. }}
 	\label{log_sigma_vs_delta_t.pdf}
 \end{figure}
As will be explained in more detail in \cite{preparation}, the factor of 2 between the two time scales can be understood from the fact that, due to transformation properties under rotations, the metric component corresponding to momentum appears squared in the computation of the time-dependent conductivity.

\end{document}